\documentclass[aps,prl,reprint,superscriptaddress,amsmath]{revtex4-1}
\usepackage{graphicx}
\usepackage{bbold}
\usepackage{amsmath}

\newcommand{\processfidelity}{$(92 \pm 2)\%$} %mc process fidelity
\newcommand{\bra}[1]{\ensuremath{\langle #1|\,}}
\newcommand{\ket}[1]{\ensuremath{\,|#1\rangle}}
\newcommand{\Ca}{\ifmmode ^{40}\text{Ca}^{+} \else $^{40}$Ca$^{+}$~\fi}
\newcommand{\mus}{\ifmmode \mu\mathrm{s} \xspace \else $\mu$s \xspace\fi}

\begin{document}

%\preprint{}

\title{Quantum-state transfer from an ion to a photon} %Title of paper



\author{A. Stute}
%\thanks{}
\thanks{These authors contributed equally to this work.}
\affiliation{Institut f{\"u}r Experimentalphysik, Universit{\"a}t Innsbruck, Technikerstra{\ss}e 25, 6020 Innsbruck, Austria}
\author{B. Casabone}
\thanks{These authors contributed equally to this work.}
\affiliation{Institut f{\"u}r Experimentalphysik, Universit{\"a}t Innsbruck, Technikerstra{\ss}e 25, 6020 Innsbruck, Austria}
\author{B. Brandst\"atter}
\affiliation{Institut f{\"u}r Experimentalphysik, Universit{\"a}t Innsbruck, Technikerstra{\ss}e 25, 6020 Innsbruck, Austria}
\author{K. Friebe}
\affiliation{Institut f{\"u}r Experimentalphysik, Universit{\"a}t Innsbruck, Technikerstra{\ss}e 25, 6020 Innsbruck, Austria}
\author{T.~E. Northup}
\affiliation{Institut f{\"u}r Experimentalphysik, Universit{\"a}t Innsbruck, Technikerstra{\ss}e 25, 6020 Innsbruck, Austria}
\author{R.Blatt}
\affiliation{Institut f{\"u}r Experimentalphysik, Universit{\"a}t Innsbruck, Technikerstra{\ss}e 25, 6020 Innsbruck, Austria}
\affiliation{\"Osterreichischen Akademie der Wissenschaften, Technikerstra{\ss}e 21a, 6020 Innsbruck, Austria}
%\ $*$ These authors contributed equally to this work}


\date{\today}

\begin{abstract}
A quantum network \cite{Kimble08a,Duan10} requires information transfer between distant quantum computers, which would enable distributed quantum information processing \cite{Cirac99,Barrett05,Lim05} and quantum communication \cite{Briegel98,DiVincenzo00}.
One model for such a network is based on the probabilistic measurement of two photons, each entangled with a distant atom or atomic ensemble, where the atoms represent quantum computing nodes \cite{Duan01,Browne03,Chou07,Moehring07,Hofmann12}.
A second, deterministic model transfers information directly from a first atom onto a cavity photon, which carries it over an optical channel to a second atom \cite{Cirac97}; a prototype with neutral atoms has recently been demonstrated \cite{Ritter12}.
In both cases, the central challenge is to find an efficient transfer process that preserves the coherence of the quantum state.
Here, following the second scheme, we map the quantum state of a single ion onto a single photon within an optical cavity.
Using an ion allows us to prepare the initial quantum state in a deterministic way \cite{Leibfried03,Haeffner08}, while the cavity enables high-efficiency photon generation \cite{McKeever04a, Keller04, Hijlkema07, Barros09}.  The mapping process is time-independent, allowing us to characterize the interplay between efficiency and fidelity.  As the techniques for coherent manipulation and storage of multiple ions at a single quantum node are well established \cite{Leibfried03,Haeffner08}, this process offers a promising route toward networks between ion-based quantum computers.
\end{abstract}


\maketitle %\maketitle must follow title, authors, abstract and \pacs

% References should be done using the  \cite, \ref, and \label commands
%\section{}
%\label{}
%\subsection{}
%\subsubsection{}

% If in two-column mode, this environment will change to single-column format so that long equations can be displayed. 
%\begin{widetext}
%$$\mbox{put long equation here}$$
%\end{widetext}

% Figures should be put into the text as floats. 
% Use the graphics or graphicx packages (distributed with LaTeX2e).
% Put the label that you will use with \ref{} command in the braces of the \label{} command.
%
% \begin{figure}
% \end{figure}

% \begin{table}
% \end{tabular}
% \end{table}

In the original proposal for quantum-state transfer \cite{Cirac97}, a photonic qubit comprises the number states $\ket{0}$ and $\ket{1}$. Such a qubit was subsequently employed for the cavity-based mapping of a coherent state onto an atom \cite{Boozer07a}.  However, due to losses in a realistic optical path, it is advantageous instead to encode the qubit within a degree of freedom of a single photon.  As a frequency qubit \cite{Olmschenk09} would be challenging to realize reversibly within a cavity, we choose the polarization degree of freedom. The target process then maps an electronic superposition of atomic states $\ket{S}$ and $\ket{S'}$ to the polarization state $\ket{H}$ and $\ket{V}$ of a photon,
\begin{align}
	\left( \cos\alpha \ket{S} + e^{i \varphi} \sin\alpha \ket{S'} \right) \otimes \ket{0} \longrightarrow  \notag \\
	\ket{D} \otimes \left( \cos\alpha \ket{H} + e^{i \varphi} \sin\alpha \ket{V} \right),
\label{eq_mapping}
\end{align}
preserving the superposition's phase and amplitude, defined by $\varphi$ and $\alpha$; $\ket{D}$ is a third atomic state.

\begin{figure}
\includegraphics{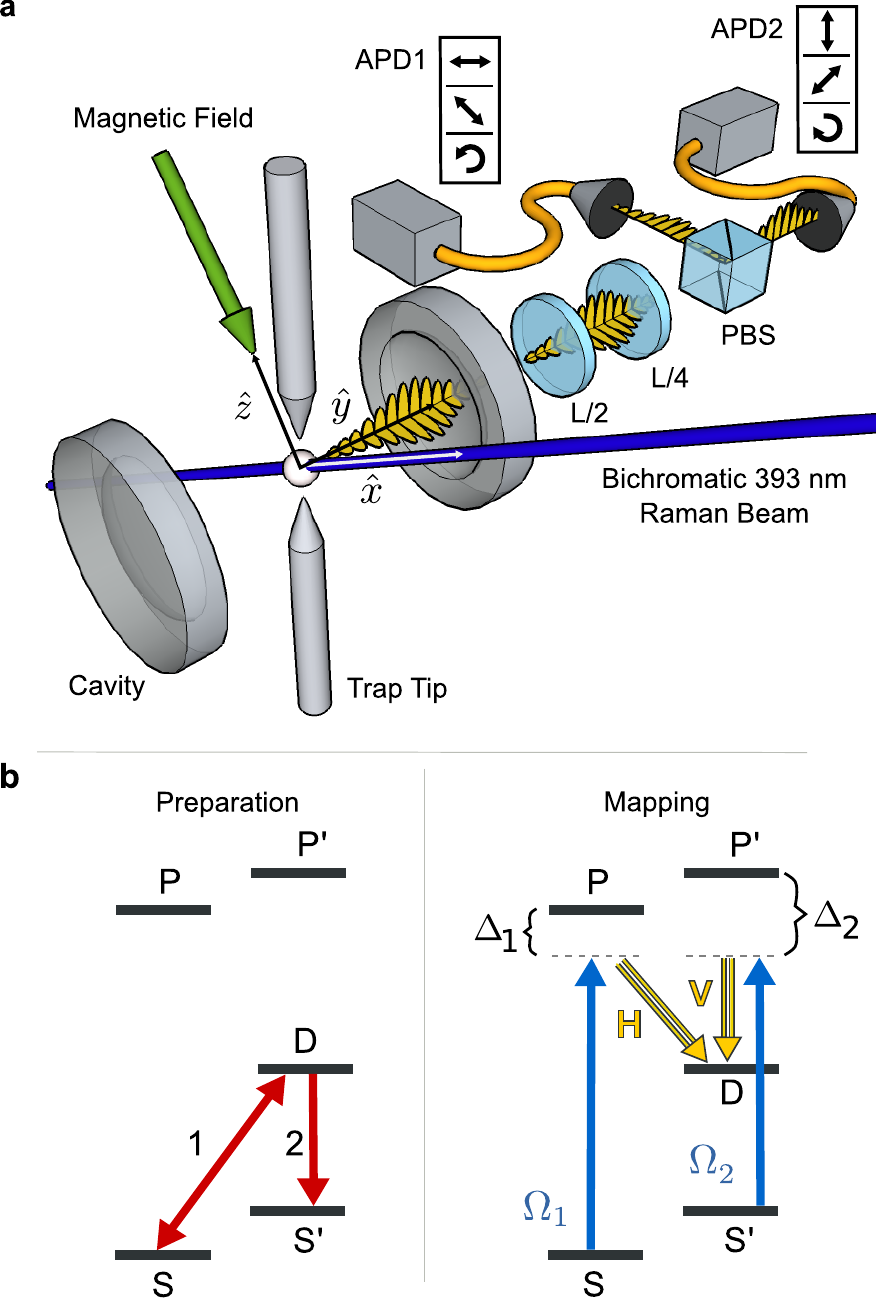}
\caption{\label{setup} \textbf{Experimental configuration and mapping sequence.}
\textbf{a}, A \Ca ion is confined in a linear Paul trap (indicated schematically by two trap tips) and positioned at the antinode of an optical cavity.  A bichromatic 393 nm field drives a pair of Raman transitions, generating a single cavity photon.
The beam is linearly polarized in the $\hat z$ direction and propagates along $\hat x$. An external magnetic field is chosen parallel to $\hat z$, orthogonal to the cavity axis $\hat y$, in order to drive $\pi$ transitions.
% The beam is polarized parallel to an external magnetic field and orthogonal to the cavity axis in order to drive $\pi$ transitions.
The measurement basis of photons exiting the cavity is set by half- and quarter-waveplates (L/2, L/4).
Photons are then separated by a polarizing beamsplitter (PBS) for detection on avalanche photodiodes (APD1, APD2).
Tables indicate the polarization of photons at APD1 and APD2 corresponding to three measurement bases.
% $H/V$ and $D/A$ and one circular polarization basis $R/L$.
\textbf{b}, Two laser pulses at 729~nm (1,2) prepare the ion in a superposition of levels $S$ and $S'$.  This superposition is subsequently mapped onto the vertical $(V)$ and horizontal $(H)$ polarization of a cavity photon.
%The 393 nm laser couples $S$ and $S'$ at Rabi frequencies $\Omega_1$ and $\Omega_2$ to virtual levels that are detuned from levels $P$ and $P'$ by $\Delta_1$ and $\Delta_2$ (approximately 400~MHz).  The cavity field then couples the virtual levels to the metastable $D$ level.}
The 393 nm laser and the cavity field couple $S$ and $S'$ to the metastable $D$ level. The bichromatic laser field is detuned from levels $P$ and $P'$ by $\Delta_1$ and $\Delta_2$ (approximately 400~MHz) and has Rabi frequencies $\Omega_1$ and $\Omega_2$. % The cavity field then couples the virtual levels to the metastable $D$ level.
}
\end{figure}

As an atomic qubit, we use two electronic states of a single \Ca ion in a linear Paul trap within an optical cavity \cite{Stute11} (Fig.~\ref{setup}a).
%(see  \cite{Stute11} for further details of the experimental apparatus)
Any superposition state of the atomic qubit can be deterministically initialized %with high efficiency 
via coherent laser manipulations \cite{Leibfried03,Haeffner08}, where this initialization is independent of the ion's interaction with the cavity field.
%This particularly means that control over the atomic qubit is given completely .
Following optical pumping to the Zeeman state $\ket{S} \equiv \ket{4^{2}S_{1/2}, m_J = - 1/2}$, the atomic qubit is encoded in the states $\ket{S}$  and $\ket{S'} \equiv \ket{4^{2}S_{1/2}, m_J = + 1/2}$
via two laser pulses on the quadrupole transition that couples the $4^{2}S_{1/2}$ and $3^{2}D_{5/2}$ manifolds (Fig.~\ref{setup}b).   The length and phase of a first pulse on the $\ket{S} \leftrightarrow \ket{D} \equiv \ket{3^{2}D_{5/2}, m_J = +1/2}$  transition set the amplitude and phase of the initial state.  The state is subsequently transferred back to the $S$ manifold via a $\pi$-pulse on the $\ket{D} \leftrightarrow \ket{S'}$ transition.

To implement the state-mapping process of Eq.~\ref{eq_mapping}, we drive two simultaneous Raman transitions in which both states $\ket{S}$ and $\ket{S'}$ are coupled to the same final state $\ket{D}$ via intermediate states $\ket{P}$ and $\ket{P'}$ (Fig.~\ref{setup}b).  One arm of both Raman transitions is driven by a laser, the second arm is mediated by the cavity field, and a single photon is generated in the process \cite{McKeever04a, Keller04, Hijlkema07, Barros09}. If the initial state was $\ket{S}$, this photon is in a horizontally polarized state $\ket{H}$; if it was $\ket{S'}$, a vertically polarized photon $\ket{V}$ is generated.
As the polarization modes of the cavity are degenerate, entanglement of the polarization with the frequency degree of freedom is avoided.
We have recently used a similar Raman process to generate ion--photon entanglement \cite{Stute12}.  In contrast, here, by coupling two initial atomic states to one final state, the ion's electronic state is transferred coherently to the photon, and no information remains in the ion.  The crux of this mapping problem is to maintain amplitude and phase relationships during the transfer process.

%\textbf{Bichromatic Raman transition:} \\
A magnetic field of 4.5~G lifts the degeneracy of electronic states $\ket{S}$ and $\ket{S'}$, so that
the two Raman transitions have different resonance frequencies. We therefore apply a phase-stable bichromatic driving field with detunings $\Delta_1$ and $\Delta_2$ from $\ket{P}$ and $\ket{P'}$, respectively (Fig.~\ref{setup}b). If the difference frequency of the bichromatic components is equal to the energy splitting of the qubit states $\ket{S}$ and $\ket{S'}$, both Raman transitions are driven resonantly.
%is mapped to the polarization of a single photon $\ket{H}, \ket{V}$ via a bichromatic Raman field.
The Rabi frequencies of the transitions are determined not only by the field amplitudes $\Omega_1$ and $\Omega_2$ but also by atomic transition probabilities and by the cavity orientation with respect to the magnetic field \cite{Stute11}.  In order to preserve the amplitudes of the initial state during mapping, we balance the Raman transition probabilities to compensate for these factors by setting $\Omega_1=2\Omega_2$.

%\textbf{Description of the tomograpic measurements:} \\
The mapping process is characterized via process tomography, in which the bichromatic Raman transition is applied to four orthogonal initial states of the atom: $\ket{S}, \ket{S'}, \ket{S-S'}, \ket{S+iS'}$.
For each input state, we measure the polarization state of the output photon via state tomography, using three orthogonal measurement settings \cite{James01} selected with two waveplates before a polarizing beamsplitter (Fig.~\ref{setup}a).  Avalanche photodiodes detect photons at both beamsplitter output ports.
%(time resolution?)
%The atomic qubit is initialized independently of the cavity mode via coherent manipulation on the quadrupole transition, a deterministic and efficienct process (efficiency - fidelity: measure ?).
%The photon tomography is measured with two waveplates and a PBS.

\begin{figure*}
\includegraphics{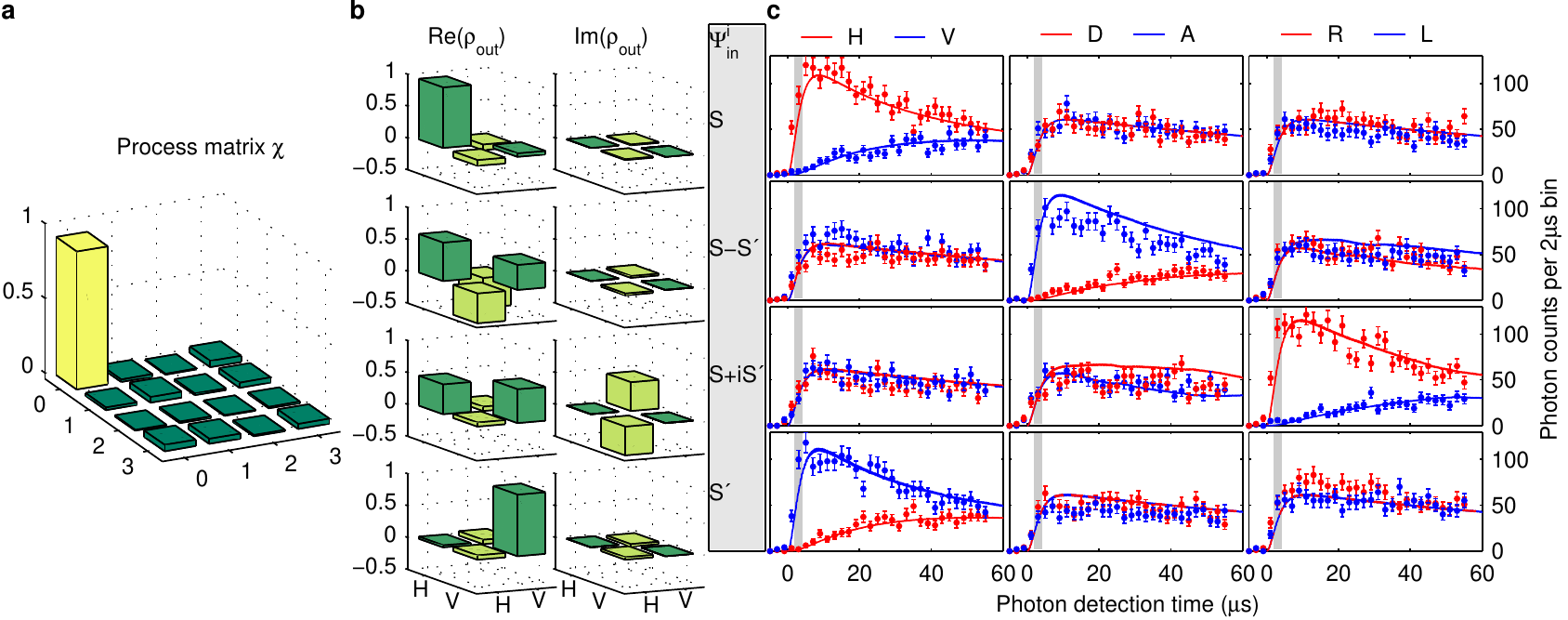}
\caption{\label{fig_tomography} \textbf{Process and state fidelities of the ion--photon mapping.} \textbf{a}, Absolute values of the process matrix $\chi$ reconstructed from cavity photons detected between 2~$\mu$s and 4~$\mu$s after the bichromatic field is switched on at time~$=0$. \textbf{b}, State tomography of photons in the same time window for the four input states \ket{S},\ket{S-S'},\ket{S+iS'}, and \ket{S'}, shown in rows from top to bottom.  \textbf{c}, Each state tomography corresponds to measurements in the three bases $H/V$, $D/A$, and $R/L$ (columns). For each input state, the temporal shapes of single photons are plotted in red for polarizations $H,D,R$ and in blue for polarizations $V,A,L$.  In each row, in two of three columns,
%In two of three measurement bases,
photons are equally distributed over both detectors, while in the third, photons are generated ideally with a single polarization.  Master-equation simulations (red and blue lines) successfully reproduce the observed dynamics.  The grey shaded area indicates the time window used for tomography.
%Each row corresponds to one input state, each column to one polarization basis.
}
\end{figure*}

Process tomography extracts the process matrix $\chi$, which parameterizes the map from an arbitrary input density matrix $\rho_\mathrm{in}$  to its corresponding output state $\rho_\mathrm{out}$ in the basis of the Pauli operators $\sigma_{0,1,2,3} \equiv \{\mathbb{1}, \sigma_x, \sigma_y, \sigma_z \}$:  $\rho_\mathrm{out} = \sum_{i,j} \limits \chi_{i,j} \sigma_i \rho_\mathrm{in} \sigma_j $. As the ideal mapping process preserves the qubit, the overlap $\chi_{0,0}$ with the identity should be equal to one. We identify $\chi_{0,0}$ as the process fidelity, which quantifies the success of the mapping. A maximum likelihood reconstruction \cite{Jezek03} of $\chi$ is plotted in Fig.~\ref{fig_tomography}a for a 2~$\mu$s window of photons exiting the cavity.  Here the matrix element $\chi_{0,0}$ indicates a process fidelity  of \processfidelity, well above the classical threshold of 1/2.
%After the identity component, the $(I,Z)$ terms are the next largest in the process matrix, indicating dephasing as a primary source of error.
Other diagonal elements $\chi_{1,1}=(3\pm1)\%$ and $\chi_{3,3}=(4\pm2)\%$ reveal a minor depolarization of the quantum state.

Another metric for quantum processes is the mean state fidelity, which evaluates the state fidelities $\bra{\psi_\mathrm{in}^i} \rho_\mathrm{out}^i \ket{\psi_\mathrm{in}^i}$ for a set of input states $\ket{\psi_\mathrm{in}^i}$, where $\rho_\mathrm{out}^i$ represent the corresponding photon output states.  The mean state fidelity can also be directly extracted from the process fidelity for an ideal unitary process \cite{Horodecki99}.  For each of our four input states, state tomography of the output photon is shown in Fig.~\ref{fig_tomography}b, using the same photon collection window as in Fig.~\ref{fig_tomography}a.  The corresponding state fidelities are $(96\pm1)\%$ for $\ket{S}$, $(94\pm2)\%$ for $\ket{S'}$, $(97\pm2)\%$ for $\ket{S-S'}$, and $(95\pm2)\%$ for $\ket{S+iS'}$, yielding a mean of $(96\pm1)\%$.
This agrees with the value of $(95\pm1)\%$ extracted from the process fidelity and exceeds the classical threshold \cite{Horodecki99} of 2/3.

%\textbf{Results B: Photon wavepackets} (more than one paragraph)\\
We now consider the evolution over time of the photonic output states $\rho_\mathrm{out}^i$ generated from these four atomic input states.  In Fig.~\ref{fig_tomography}c, we plot the temporal shape of the emitted photon in each of three measurement bases, a total of 12 cases.  For each input state, there exists one polarization measurement basis in which photons would ideally impinge on only one detector.  If the ion is prepared in the state \ket{S} and measured in the $H/V$ polarization basis, for example, the mapping scheme of Fig.~\ref{setup}b should only produce the photon state \ket{H}.  However, a few microseconds after the Raman driving field is switched on, we see that the photon state \ket{V} appears and is generated with increasing probability over the next 55~$\mu$s.  The mechanism here is off-resonant excitation of the $4^{2}P_{3/2}$ manifold and decay to the previously unpopulated state \ket{S'}, followed by a Raman transition generating the `wrong' polarization.  
If the ion is prepared in \ket{S'}, the temporal photon shapes are inverted and symmetric, with the initial state \ket{V} followed by the gradual emergence of \ket{H}.
We have confirmed this process through master-equation simulations of the ion--cavity system, also plotted in Fig.~\ref{fig_tomography}c and described in the Supplementary Information.

For the superposition input states \ket{S-S'} and \ket{S+iS'}, the mapping generates a photon with antidiagonal polarization $A=(H-V)/\sqrt{2}$ and right-circular polarization $R=(H+iV)/\sqrt{2}$, respectively. Thus, photons impinge predominantly on one detector in the 
%Thus, measurements in the the measurement bases in which a single photon polarization is generated are 
diagonal(D)/antidiagonal(A) and right(R)/left(L) bases, where $D=(H+V)/\sqrt{2}$ and $L=(H-iV)/\sqrt{2}$ (Fig.~\ref{fig_tomography}c).
% (see Supplementary Information).
Here, as for states \ket{S} and \ket{S'}, photons with the `wrong' polarization are due to off-resonant scattering before the mapping occurs. In this case, scattering destroys the phase relationship between the $S$ and $S'$ components.
%leads to dephasing rather than population of an empty state, but with the same result that coherence is reduced.
(Note that for eight of the cases in Fig.~\ref{fig_tomography}c, the measurement basis projects the photon polarization onto the two detection paths with equal probability.)

\begin{figure}
\includegraphics{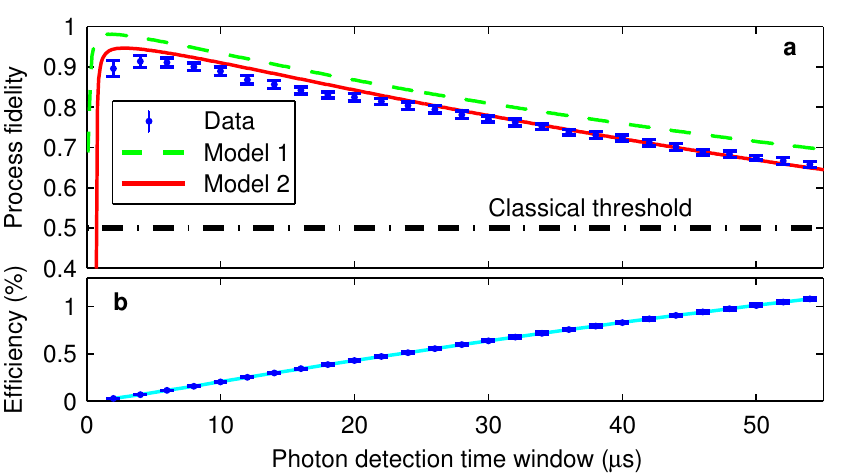}
\caption{\label{fig_efficiency} \textbf{Time dependence of process fidelity and efficiency.} \textbf{a}, Cumulative process fidelity and \textbf{b}, process efficiency are plotted as a function of the photon-detection time window, where error bars represent one s.d. (See Supplementary Information.)  A green dashed line indicates the simulated process fidelity for the same parameters as in Fig. \ref{fig_tomography}.
%, representing fidelities achievable with our experimental apparatus.
To this model, we now add the effects of detector dark counts, imperfect state initialization, and magnetic-field fluctuations, quantified in independent measurements, with the result indicated by a red line. A fit to the process efficiency is used to weight the effect of dark counts. The second model agrees well with the data, while the first one represents achievable values for this ion--cavity system.} 
\end{figure}

The accumulation of scattering events over time suggests that the best mapping fidelities can be achieved by taking into account only photons detected within a certain time window.  Such a window is used for the preceding analysis of process and state fidelities.
%which includes photons detected between 7 and 9~$\mu$s after the bichromatic driving field is switched on.
%\textbf{Results C: Efficiency + time independence}\\
For each attempt to prepare and map the ion's state, the probability to detect a photon within this window is $4 \cdot 10^{-4}$, which we identify as the process efficiency.
%The overall efficiency of the process within this window is $4 \cdot 10^{-4}$, including qubit initialization, mapping and photon detection.
This efficiency can be increased at the expense of fidelity by considering a broader time window.
%of the How can we anyways integrate over more than 100~ns? Larmor precession is counteracted by the frequency difference of the two Raman fields, resulting in time-independence (see methods). This is great: We can thus read out our stationary qubit at any time although the qubit states are non-degenerate.
%We can thus always take into account all photons.
Fig.~\ref{fig_efficiency} shows both the cumulative process fidelity and efficiency as a function of the photon-detection window.
%During the first x~$\mu$s,
%The fidelity reaches a maximum after 9~$\mu$s.
The fidelity initially increases because at short times ($< 100$ ns), photons are produced primarily via the off-resonant rather than the resonant component of the Raman process and thus are not in the target polarization state.
This coherent effect, which we have investigated through simulations, is quickly damped due to the low amplitude of off-resonant Raman transitions.
%with respect to off-resonant Raman transitions ($\delta/(2\pi) \approx 5$~MHz).
The cumulative process fidelity reaches a maximum between 2~$\mu$s and 4~$\mu$s after the bichromatic driving field is switched on, the time interval used to analyze the data of Fig.~\ref{fig_tomography}a and b.  The fidelity then slowly decreases as a function of time due to the increased likelihood of off-resonant scattering. 

If all photons detected within 55~$\mu$s are taken into account, the process efficiency exceeds 1\%, while the process fidelity of $(66\pm1\%)$ remains above the classical threshold of 1/2. % of 50\%.
This process efficiency includes losses in the cavity mirrors, output path, and detectors. The corresponding probability for state transfer within the cavity is 14.7\%. 
A longer detection time window would allow transfer probabilities approaching one, but fidelities would fall below the classical threshold.
Simulations that include the effects of detector dark counts, imperfect state initialization, and magnetic-field fluctuations agree well with the data of Fig.~\ref{fig_efficiency}.  In the absence of these three effects, simulations indicate that fidelities of 98\% would be possible in our ion--cavity system.

The atomic superposition of $\ket{S}$ and $\ket{S'}$ experiences a 12.6~MHz Larmor precession, which corresponds to a rotation of the states' relative phase at this frequency. One might expect that as a result, it would not be possible to bin data from photons generated from this superposition across a range of arrival times as described above. However, because the frequency difference $\Delta_1-\Delta_2$ of the bichromatic Raman field matches the frequency difference between the two states, the Raman process generates a photon that preserves the initial states' relative phase.  As a result, the phase of the photon superposition is independent of detection time.  (See Supplementary Information.) This transfer scheme thus offers advantages for any quantum system in which a magnetic field lifts the degeneracy of the states encoding a qubit.

%\textbf{Summary:} \\
Following the deterministic initialization of an atomic qubit within a cavity, we have shown the coherent mapping of its quantum state onto a single photon. %Employing a bichromatic Raman scheme, the nondegenerate qubit can be mapped to the photon at any time.
The mapping scheme achieves a high process fidelity, and by accepting compromises in fidelity, we increase the efficiency of the process within the cavity up to 14\%.
%\textbf{Outlook + applications:} \\
% Error budget
The transfer measurement is primarily limited by detector dark counts at 5.6~Hz, imperfect state initialization with a fidelity of 99\%, magnetic-field fluctuations corresponding to an atomic coherence time of 110~$\mu$s, and the finite strength of the ion--cavity coupling in comparison to spontaneous decay rates.
While a stronger coupling would improve the fidelity for a given efficiency, we note that the mapping fidelity in our current intermediate-coupling regime could also be improved by encoding the stationary qubit across multiple ions \cite{Lamata11}.
%\textbf{Possible applications:} \\
A direct application of this bichromatic mapping scheme is state transfer between two remote quantum nodes \cite{Cirac97,Ritter12}. Furthermore, via a modified bichromatic scheme, a single ion-cavity system can act as a deterministic source of photonic cluster states \cite{Lindner09}, an essential resource for measurement-based quantum computation \cite{Raussendorf01}.

We thank T. Monz and P. Schindler for assistance in tomography analysis and P. O. Schmidt for early contributions to the experiment design.
This work was supported by the Austrian Science Fund (FWF), the European Commission (AQUTE), the Institut f\"ur Quanteninformation GmbH,
and a Marie Curie International Incoming Fellowship within the 7th European Framework Program.

\section*{Supplementary Information}
\subsection{Coupling parameters} 
The coherent ion-cavity coupling rate is $g = 2 \pi \times 1.4$~MHz. The cavity field decay rate is $\kappa = 2 \pi \times 0.05$~MHz, and the atomic polarization decay rate from the  $P_{3/2}$ state is $\gamma = 2 \pi \times 11.5$~MHz.
The effective coupling strength of the two Raman transitions $i=1,2$ is given by $\Omega^\mathrm{eff}_i = 2 \pi \times \frac{g G_i \Omega_i}{2 \Delta_i}$, where $G_i$ is a geometric factor that takes into account both the relevant Clebsch-Gordon coefficients and the projection of the vacuum-mode polarization onto the atomic dipole moment \cite{Stute11}.
The detunings $\Delta_{1,2}$ are approximately 400~MHz.
We ensure equal transition probabilities for the two Raman transitions by setting the ratio of the Rabi frequencies $\Omega_1/\Omega_2 = G_2/G_1 = 2$, with absolute values $\Omega_1=2 \pi \times 17.5$~MHz and $\Omega_2=2 \pi \times 8.75$~MHz. %For these values, the effective coupling rate $\Omega_\mathrm{eff} \approx 2 \pi \times 16$~kHz is larger than
%the effective atomic decay rate $\gamma_\mathrm{eff} = \left(\frac{\Omega_1+\Omega_2}{2\Delta_{1,2}}\right)^2 \cdot \gamma \approx 2 \pi \times 13$~kHz, where $\Delta_{1,2} \approx 400$~MHz.
%{\color{blue} Bernardo is checking the numbers in this paragraph!}

\subsection{Process tomography and measurement bases} 
In order to characterize the mapping process, we carry out process tomography.
For this purpose, we carry out state tomography of the photonic output state for four orthogonal atomic input states \cite{Nielsen2000}.
Each state tomography of a single photonic polarization qubit consists of measurements in the three bases $H/V$, $D/A$, and $R/L$.
Note that $H$ and $V$ correspond to the two cavity modes, where $V$ is parallel to the $\hat{z}$ axis of Fig.~1a in the main text and $H$ to the $\hat{x}$ axis.  (In contrast, in Ref. \cite{Stute12}, we defined $H$ rather than $V$ to be parallel to the magnetic field axis.)
%Along with the $H/V$ basis, we define two additional bases $D/A$ and $R/L$, with diagonal, antidiagonal, right, and left basis vectors
%\begin{align*}
%D &= (H+V)/\sqrt{2}, \\
%A &= (H-V)/\sqrt{2}, \\
%R &= (H+iV)/\sqrt{2}, \\
%L &= (H-iV)/\sqrt{2}. \\
%\end{align*}
In each basis, we perform two measurements of equal duration in which the output paths to avalanche photodiodes APD1 and APD2 are swapped by rotating waveplates L/2 and L/4.  Summing these measurements allows us to compensate for unequal detection efficiencies in the two paths \cite{Stute12}.

The process and density matrices plotted in Fig.~2a and b in the main text are reconstructed from the data using a maximum likelihood fit  \cite{Jezek03}.  The data consists of $32,400$ single-photon detection events.
We extract fidelities and their statistical uncertainties via non-parametric bootstrapping assuming a multinomial distribution \cite{Efron93}.  Statistical uncertainties are stated as one standard deviation.
%In Fig.~\ref{fig_efficiency} we evaluate process tomography as a function of photon detection time.  (Say something about bin size in text?)
%\vspace{.4cm} \\

\subsection{Time independence} 
If the atomic qubit is comprised of nondegenerate states, Larmor precession will change the qubit's phase over time.
For a monochromatic mapping protocol in which the photonic qubit is encoded solely in polarization, the phase of the photonic qubit thus depends on the time of photon generation \cite{Specht11}.
In contrast, for the two Raman fields $\Omega_1 e^{i \omega_{l_1} t}$ and $\Omega_2 e^{i \omega_{l_2} t}$ at frequency $\omega_{l_1}$ and $\omega_{l_2}$, the mapping pulse can be applied at any time for the correct choice of frequency difference between the two Raman fields $\omega_{l_1}-\omega_{l_2} = \Delta E_{S,S'}/\hbar$, where $\Delta E_{S,S'}$ is the Zeeman splitting between the two qubit states $S$ and $S'$.
In this case, the atomic qubit is always mapped to the same photonic qubit state, independent of photon generation time.
%In this case, the phase of the photon state is constant although the atomic state undergoes Larmor precession.
%This is similar as for the recently reported bichromatic entanglement protocol \cite{Stute12}, where the phase of the atomic state after photon detection remains time-independent.

To show this, we define a model system consisting of initial states $\ket{S,n}, \ket{S',n}$, intermediate states $\ket{P,n}, \ket{P',n}$ and target state $\ket{D,n}$ with energies $E_{S,S',P,P',D} = \hbar \omega_{S,S',P,P',D}$. Here, $n=0,1$ denotes the number of photons in either of the two degenerate cavity modes at energy $\hbar \omega_C$. A similar model system was used to explain the time independence of the bichromatic entanglement protocol that we recently demonstrated \cite{Stute12}. The $\ket{S,n}\leftrightarrow\ket{P,n}$ transition is driven by the field $\Omega_1 e^{i \omega_{l_1} t}$ with detuning $\Delta_{l_1} = \omega_S - \omega_P - \omega_{l_1}$, while the $\ket{S',n}\leftrightarrow\ket{P',n}$ transition is driven by the field $\Omega_2 e^{i \omega_{l_2} t}$ with detuning $\Delta_{l_2} = \omega_{S'} - \omega_{P'} - \omega_{l_2}$.
We choose a unitary transformation into a rotating frame that takes into account the atomic precession at frequency $\omega_S-\omega_{S'}$:
$U=e^{-i\omega_{l_1}t\ket{S}\bra{S}} ~ e^{-i\omega_{l_2}t\ket{S'}\bra{S'}}.$
After this transformation and adiabatic elimination of the state $\ket{P,n}$, the Hamiltonian reads
\begin{align}
\mathcal{H}/\hbar = & (\omega_S + \omega_{l_1}) \ket{S}\bra{S} + (\omega_{S'}  + \omega_{l_2}) \ket{S'}\bra{S'} \nonumber \\
&+ \omega_{P'}  \ket{P'}\bra{P'} + \omega_{D}  \ket{D}\bra{D} + \omega_C  \ket{1}\bra{1} \nonumber \\
&+ \left(g^{\mathrm{eff}}_1  \ket{D,1}\bra{S,0} + g^{\mathrm{eff}}_2 \ket{D,1}\bra{S',0}
+ \mathrm{h.c.} \right) ,
\end{align}
where the energy reference is the $\ket{P,0}$ state ($\omega_P=0$). Both couplings
$g^{\mathrm{eff}}_i = \frac{\Omega_i \cdot g}{2 \Delta_{l_i}}$
are time-independent.
%and therefore the constant offset $\hbar \omega_P \1$ has been added to $\mathcal{H}$, and
Choosing the frequencies of the two fields to match the two Raman conditions \\ $\omega_S+\omega_{l_1} = \omega_D+\omega_C = \omega_{S'}+\omega_{l_2}$ corresponds to a frequency difference $|\omega_{l_1}-\omega_{l_2}| = |\omega_S-\omega_{S'}|$.
The two states $\ket{S,0}$ and $\ket{S',0}$ are degenerate in this frame,
%we calculate the energy of $\ket{S',0}$ to be
%\begin{equation}
%\omega_{S'} + \omega_{l_2} = \omega_S - (\omega_S - \omega_{S'}) + \omega_{l_1} - ( \omega_{l_1} - \omega_{l_2}) = \omega_{S} + \omega_{l_1}.
%\end{equation}
%We conclude that the states $\ket{S,0}$ and $\ket{S',0}$ are degenerate in this frame
resulting in a constant phase of the atomic state. As the couplings are also time-independent, the phase $\varphi$ of the atomic state does not change during the transfer to the photonic state (equation~1 of the main text). As both modes of the cavity are degenerate, the phase $\varphi$ of the photonic state remains constant after the transfer.

So far, we have neglected off-resonant Raman transitions, i.e., $\Omega_{1}$ coupling $\ket{S',n}$ to $\ket{P',n}$ and $\Omega_{2}$ coupling $\ket{S,n}$ to $\ket{P,n}$. Taking these couplings into account, the terms $g^{\mathrm{eff}}_i$ in the Hamiltonian  are proportional to
$\Omega_i + \Omega_j e^{i (\omega_{l_j}-\omega_{l_i}) t}$ after transformation into the rotating frame and adiabatic elimination of $\ket{P,n}$.
Here, the second term, oscillating at $|\omega_{l_i}-\omega_{l_j}|$, corresponds to off-resonant Raman transitions in which a photon with unwanted polarization is generated. These terms are neglected in the rotating wave approximation because $|g^{\mathrm{eff}}_i| \ll |\omega_{l_1}-\omega_{l_2}|$.
These off-resonant coupling terms, however, explain why the fidelity of the mapping process only reaches its maximum after about 3~$\mu$s (Fig.~3 of the main text). As confirmed by our simulations, the off-resonant Raman processes generate photons with unwanted polarization at the timescale of 100~ns after turning on the drive laser pulse. However, the probability for this process is very low due to the large detuning from Raman resonance, and the effect is quickly overcome by the much higher probability of generating photons with the desired polarization thereafter.
%The low probability explain why the very first photons are produced with wrong polarization of the first photons starts at intermediate values in the analysis of the process fidelity over integration time (see below).

\subsection{Simulations} 
Numerical simulations of the state-mapping process are based on the Quantum Optics and Computation Toolbox for MATLAB \cite{Tan99}. We formulate the master equation for the $18$-level $^{40}$Ca$^+$ system interacting with two orthogonal modes of an optical cavity. We then numerically integrate the master equation to obtain the system's density matrix as a function of time. The simulation includes atomic and cavity decay and the laser linewidth.  Relative motion of the ion with respect to the cavity mode is taken into account by introducing an effective atom-cavity coupling $g_{\mathrm{motion}}$ smaller than $g$. This motion results from the (presumably mechanical) oscillation of the ion trap with respect to the cavity \cite{Stute11}. Furthermore, small effects such as finite switching time of the laser, laser-amplitude noise and relative phase noise are neglected in the model.

The simulations require us to specify the input parameters: magnetic field $B$, Raman-laser frequencies $\omega_{l_1}$ and $\omega_{l_2}$, photon detection path efficiency, and Rabi frequencies $\Omega_1$ and $\Omega_2$, as well as system parameters $g$, $\kappa$, and $\gamma$.  $B$ and the laser frequencies are determined from spectroscopy of the quadrupole transition to within $3$~kHz. A detection path efficiency of $6.8\%$ is used to scale the simulation results, consistent with previous measurements \cite{Stute11}. $\Omega_1$ and $\Omega_2$ are determined experimentally via Stark-shift measurements with an uncertainty on the order of 20\%. However, the temporal shape of the photons is highly dependent on $\Omega_1$ and $\Omega_2$ and on the atom-cavity coupling $g$.  In the simulation, we therefore adjust $\Omega_1$ and $\Omega_2$ within the experimental uncertainty range and find that the values $\Omega_1=2 \pi \times 17.5$~MHz and $\Omega_2=2 \pi \times 8.75$~MHz generate photon shapes that have the best agreement with data. In order to improve this agreement, we adjust $g$ to the effective value $g_{\mathrm{motion}}=0.63g$, consistent with earlier measurements \cite{Stute11}.

As discussed in the main text and presented in Fig.~2c, there are eight combinations of initial state and detection basis for which the temporal photon shapes on both detectors are identical.  However, in two of these eight cases, the simulated photon shapes in the two polarization modes do not overlap perfectly with one another.  This small discrepancy occurs for the $\ket{S+iS'}$ initial state and the $D/A$ basis as well as for the $\ket{S-S'}$ state and the $R/L$ basis, and it is due to errors that accumulate during the numerical integration routine.

For each detection time window in Fig.~3 in the main text, we estimate the relative contributions of APD dark counts and data, and the simulated density matrices are weighted accordingly.  Additionally, off-diagonal matrix terms are scaled by a factor of 0.99 representing imperfect state initialization and by the exponential $e^{-2t/\tau}$, where $\tau = 110~\mu$s is the atomic coherence time.

%\begin{acknowledgments}
% Put your acknowledgments here.
%\end{acknowledgments}

\bibliography{cqed_bibsonomy}

\end{document}